\documentstyle[preprint,aps,floats,psfig]{revtex}
\bibstyle{unsrt}

\tightenlines
\begin{document}
\draft
\preprint{MIT-CTP-3107}
 
\title{Dynamical behavior of spatially inhomogeneous relativistic 
$\lambda \phi^4$ quantum field theory in the Hartree approximation}
\author{Lu\'{\i}s M. A. Bettencourt$^{1,3}$, 
Karen Pao$^2$ and J. G. Sanderson$^{2}$}
\address{$^1$Theoretical Division, Los Alamos National Laboratory
MS B288, Los Alamos NM 87545}
\address{$^2$Computer and Computational Sciences Division, 
Los Alamos National Laboratory, MS B256, Los Alamos NM 87545}
\address{$^3$Center for Theoretical Physics, 
Massachusetts Institute of Technology,
Bldg 6-308, 77 Massachusetts Avenue, Cambridge MA 02139 USA.}
\date{\today}
\maketitle

\begin{abstract}
We study the dynamics of a spatially inhomogeneous 
quantum $\lambda \phi^4$ field theory in $1+1$ dimensions 
in the Hartree approximation. In particular, we investigate 
the long-time behavior of this approximation in a variety of controlled 
situations, both at zero and finite temperature.
The observed behavior is much richer than that in the spatially 
homogeneous case. 
Nevertheless, we show that the fields fail to thermalize in a 
canonical sense, as expected from analogous results in closely related 
(mean field) transport theory. 
We argue that this dynamical approximation 
is best suited as a means to study the short-time decay of 
spatially inhomogeneous fields and in the dynamics of coherent 
quasi-classical inhomogeneous configurations (e.g. solitons) 
in a background of dynamical self-consistent quantum fluctuations. 
\end{abstract}

\pacs{PACS Numbers: 03.70.+k, 05.70.Ln, 11.10.-z, 11.15.Kc \hfill}
 

\section{Introduction}
\label{secI}

In recent years much progress has been made in understanding  
the real time dynamics of {\it spatially homogeneous} quantum field theories.
This has been made possible by considering mean field approximations
({\it e.g.}, Hartree or the large-$N$ limit) to the full quantum 
dynamics of these systems. 
The advantage of such techniques is that they permit 
the approximate evolution of an arbitrarily out-of-equilibrium 
quantum field theory under many different circumstances, 
such as in the presence of large mean fields and/or at symmetry 
breaking phase transitions, where perturbative methods fail.

Numerous applications of this type of dynamics have been considered, ranging 
from symmetry breaking phase transitions \cite{SSB}, 
decoherence and dephasing \cite{decoherence},  
the formation of disoriented chiral condensates at the collisions of 
relativistic heavy ions \cite{DCCs}, 
to the re-heating and preheating of the early universe \cite{reheating}, 
to name just a few. 
Nevertheless, the approximations involved in this kind of evolution lead to 
two serious limitations, standing in the way of making the 
results apply to more general and realistic circumstances. 
The first drawback is that the long-time behavior of the system 
is incorrect due to the absence of collisional effects. 
The second is that only spatially homogeneous situations could, 
until recently \cite{fermionSmit,Smit}, be considered. 
Both limitations are not a  matter of principle; they can 
be overcome with large increases in computational effort, 
although theoretical progress, {\it e.g.}, the development of a time-local 
formulation of the quantum field dynamics including scattering, 
would be welcomed.

In this paper we lift the constraint on the spatial homogeneity 
of the system.
Mean field theories can be very naturally generalized 
to spatially inhomogeneous circumstances, allowing the study 
of the (approximate) evolution of quantum fluctuations 
in the background of dynamical, spatially dependent fields. 
This generalization results in a demanding numerical problem, 
requiring the solution of a large set of coupled {\it partial} 
differential equations. As will be made clear later, 
the size of the computational effort scales as $L^{2D}$,
where $L$ is the linear dimension of the problem
and $D$ the number of space dimensions.  Thus problems in lower spatial 
dimensions are significantly more accessible. 

One question of great general interest is whether by considering spatially 
inhomogeneous  situations one improves on the long-time properties of 
the mean field approximation. 
In particular we will investigate if the long-time regime is in 
any sense universal and, in particular,  if it corresponds to 
canonical thermal equilibrium. 
Sall\'{e}, Smit and Vink~\cite{Smit} have recently found 
that, under averaging over an ensemble of certain classes of initial 
conditions for the mean field, the statistical occupation number of the 
fluctuations may dynamically approximate a Bose-Einstein distribution over 
a limited range of times. 
Although these results are very interesting, we believe that the 
averaging over ensembles of initial mean fields is not justified 
from first principles, since the mean field is by definition already 
the quantum ensemble average one-point function.
The potential power of inhomogeneous quantum evolutions is to handle 
situations where a specific spatially dependent expectation value of the 
quantum field exists due to some perturbation of thermal equilibrium 
or of another asymptoticly stable state. 
Restrictions to particular types of initial conditions 
defeat the general purpose of the method. 
For these reasons we will adopt a more general stance here and 
investigate the main characteristics of the mean field dynamics without 
any additional ensemble averaging.

We do this in the context of a relativistic $\lambda \phi^4$ quantum field 
theory in $1+1$ dimensions. In section \ref{secII} we define the quantum 
field theory, and the mean field equations of motion. 
In section \ref{secIII} 
we make our choice of initial conditions and discuss the  
renormalization scheme necessary to make the theory independent of 
the ultraviolet cutoff.
We proceed in section \ref{secIV} to analyze the decay of a family 
of Gaussian spatial profiles. 
This allows us to investigate the dynamical response of 
the quantum fluctuation to this field background. 
We characterize general properties of the dynamics of the mean field and 
of the quantum fluctuations and pay particular attention to their long-time
limit. We then proceed in section \ref{secV}, 
in the spirit of transport theory, to study the evolution of a mean 
field small perturbation to a state of thermal equilibrium.  
Together the results of 
section \ref{secIV} and \ref{secV} allow us to determine whether 
the model thermalizes at long times in circumstances of general interest. 
Finally we present our conclusion and outlook for further applications.

\section{Theory and definitions}
\label{secII}

We study the dynamics of a relativistic 
quantum $\lambda \phi^4$ theory in $1+1$ dimensions. 
We write the Lagrangian density as
\begin{eqnarray}
{\cal L} = {1 \over 2} (\partial_\mu \phi) (\partial^\mu \phi)
- {m^2 \over 2} \phi^2 - {\lambda \over 4} \phi^4.  
\end{eqnarray} 
The field $\phi$ is taken to be real. We will work in units 
in which the speed of light $c$ and Boltzmann's constant $k_B$ are unity
$c=1, \ k_B=1$. 

There are several different ways of obtaining mean field approximations 
to the quantum field theory, by writing truncated effective actions 
(e.g. in the large $N$ limit \cite{largeNSeff}), 
by making a Gaussian variational \emph{ansatz} in the context of a 
Schr\"odinger functional approach \cite{SchroVaransatz}, or by explicitly 
expressing higher point field correlation functions in terms of the 
one- and two-point functions, thus effectively truncating the Dyson-Schwinger 
hierarchy. While these procedures do not lead to the same approximation
to the full thermo(dynamics), the resulting equations share 
much in common. Below we adopt the familiar Hartree approximation to 
the dynamical equations for the one- and two-point functions of the 
quantum field $\phi$. The type of evolution described in this paper 
generalizes to any other mean field approximation with a similar amount
of computational effort. 

Following these remarks, we separate the quantum field 
in a mean field $\varphi \equiv \langle \phi \rangle$ and a fluctuation
field  $\hat \psi$ ($\langle \hat \psi \rangle =0$), such that 
\begin{eqnarray}
\phi(x,t)= \varphi(x,t) + \hat \psi(x,t).
\end{eqnarray}
Our mean field approximation (which coincides with the familiar Hartree 
approximation) corresponds to keeping only the connected two-point 
function in $\hat \psi$. Higher point {\em connected} correlation functions of 
the fluctuations are assumed to be zero. Thus we have, e.g., 
$\langle \hat \psi^4 \rangle = 3 \langle \hat \psi^2 \rangle^2$.  
Equations of motion for the mean field and the Wightman two-point function 
$G^>(x,t;x',t') = \langle \hat \psi(x,t) \psi(x',t') \rangle$ then become
\begin{eqnarray}
&& \left[ \Box + m^2 + \lambda \varphi^2(x,t) + 3 \lambda G^>(x,t) \right]
\varphi(x,t) = 0, \label{dyneqs}  \\
&& \left[ \Box + \chi(x,t) \right] G^>(x,t;x',t') =0,  \qquad 
\chi(x,t) = m^2 + 3 \lambda \varphi^2(x,t) + 3 \lambda G^>(x,t),
\nonumber
\end{eqnarray}
with $\Box = \partial_t^2 - \partial_x^2$, $G^>(x,t;x',t') 
=\langle \hat \psi(x,t) \hat \psi(x',t') \rangle$ and $ G^>(x,t) \equiv
G^>(x=x',t=t')$.  

The relation between the (bosonic) Wightman 
functions $G^{\stackrel{<}{>}}$ and the time ordered (Feynman) two-point 
function $G_F(x,t;x',t')$ is
\begin{eqnarray}
&& G^>(x,t;x',t') = G^<(x',t';x,t), \qquad G^>(x,t;x',t')={G^<}^*(x,t;x',t'), \\
&& G_F(x,t;x',t')= \Theta(t-t')G^>(x,t;x',t') +\Theta(t'-t)G^<(x,t;x',t'),
\nonumber
\end{eqnarray} 
where $\Theta(t)$ is the usual step function. 
The advantage of working with the Wightman functions $G^{\stackrel{<}{>}}$
is that they obey homogeneous equations of motion. 

It is convenient to solve Eqs.~(\ref{dyneqs}) by decomposing 
$G^>(x,t;x',t')$ in a complete orthonormal basis of mode fields, 
which we shall denote by $\{ \psi_k(x,t) \}$. The mode index $k$ 
may be continuous or discrete. In a general context we will denote 
traces over the basis of field modes as sums over $k$.
Then we can write the quantum field $\hat \psi(x,t)$ as an expansion 
in this basis:
\begin{eqnarray}
\hat \psi(x,t) = \sum_k \left\{ a_k \psi_k(x,t) + a_k^\dagger \psi_k^*(x,t)
\right\},  
\label{expansion}
\end{eqnarray}
where $a_k,\ a_k^\dagger$ are the annihilation and creation 
operators, respectively, obeying canonical commutation relations
\begin{eqnarray}
\left[ a_k,\ a_q^\dagger \right] = \delta_{kq}, \
\left[ a_k,\ a_q \right] = \left[ a_k^\dagger,\ a_q^\dagger \right] =0.
\label{comrel}
\end{eqnarray}

Using (\ref{expansion}) and (\ref{comrel}) and 
$\langle a_k^\dagger a_q \rangle= \delta_{kq} n_B(k)$ 
we can write $G^>(x,t;x',t')$ as 
\begin{eqnarray}
&& G^>(x,t;x',t') = \sum_k \left\{ (n_B(k) + 1) \psi_k(x,t) \psi_k^*(x',t') 
+ n_B(k) \psi_k^*(x,t) \psi_k(x',t') \right\}, \nonumber 
\end{eqnarray}
where $n_B(k)$ is the Bose-Einstein thermal occupation number distribution
\begin{eqnarray}
n_B(k) = {1 \over e^{\hbar \omega_k/T} -1},
\end{eqnarray} 
with $\omega_k$ the energy eigenvalue associated with the eigenfield 
$\psi_k(x,t)$. In particular, the equal-point function 
$G^>(x,t)$ acquires the familiar form
\begin{eqnarray}
G^>(x,t) = \sum_k \left[ 2 n_B(k) + 1 \right] \psi_k(x,t) \psi_k^*(x,t).
\label{Geq}
\end{eqnarray}
The spectrum reduces to the vacuum as $T\rightarrow 0$, 
since $n_B \rightarrow 0$. 

The equation of motion for $G^>(x,t;x',t')$ can now be written 
in terms of the mode functions $\psi_k$:
\begin{eqnarray}
\left[ \Box +\chi(x,t) \right] \psi_k (x,t)=0,
\label{psi}
\end{eqnarray}
which must be solved for all $k$. Additionally $\chi(x,t)$ can 
be expressed entirely in terms of $\varphi$ and $\psi_k$ through 
Eq.~(\ref{Geq}), so that we obtain a set of self-consistent partial 
differential equations for the fields $\varphi(x,t)$ and for each of 
the $\psi_k(x,t)$. This set is formally infinite; to obtain a good 
approximation to the dynamics in the continuum we must take a very 
large number of mode fields $\psi_k(x,t)$, thus the problem becomes 
demanding in terms of computational resources.
 
In practice, the fields $\varphi(x,t), \psi_k(x,t)$ are evolved on a 
spatial lattice. Renormalizability requires that the mode fields 
$\psi_k$ must reduce to the
vacuum at large $k$ and become plane waves. This limit places a restriction
on the modes that can be resolved dynamically: for a spatial lattice of 
$N$ sites, with spacing $\triangle x$, 
modes with wavenumber $k > 2 \pi /\triangle x$ cannot be kept as dynamical fields.
Using periodic boundary conditions in space, the $n$th wavenumber is
$k_n= 2 \pi n/N \triangle x$. Thus the number of mode functions 
is restricted to $N$ per linear dimension.
Then we see that the computational effort in $D$ space dimensions 
corresponds to that of solving $N^D$ coupled partial differential 
equations on a grid of size $N^D$ points, i.e. it scales as $N^{2D}$ per 
time step, or like the $D$-dimensional volume squared. 
This scaling makes the evolution on large spatial lattices in 
three dimensions very demanding\footnote{The limits of present computational 
power allow for the solution of the three dimensional problem on a 
lattice of size $N^3$, with $N\sim O(10^2)$, for short times.}. 
For the rest of this paper we will be concerned with time evolution 
of one-dimensional fields.  

Finally, we need to specify a basis $\left\{ \psi_k(x,t) \right\}$, together 
with the scalar product 
\begin{eqnarray}	
&& \left( \psi_k(x,t) . \psi_q(x,t) \right)= i \int d x \left\{ 
\psi_k^*(x,t) \partial_t \psi_q(x,t) - \psi_q(x,t) \partial_t \psi_k^*(x,t)
\right\}. \label{sproduct}
\end{eqnarray}
It follows directly from the form of the equations of motion (\ref{dyneqs}) 
and (\ref{psi}) 
that the scalar product (\ref{sproduct}) is invariant under the dynamics, 
{\it i.e.}, $\partial_t \left( \psi_k(x,t) . \psi_q(x,t) \right)=0 $ for all $ k,q $.
This guarantees that the orthonormality of the basis, chosen at the initial 
time, remains valid to all subsequent times, although the set of fields
$\psi_k$ will evolve away from their initial conditions. 

At the initial time, it is convenient to parameterize the set of 
functions $\psi_k$ by
\begin{eqnarray}
\psi_k(x,t) = \sqrt{\hbar \over 2 \omega_k} e^{- i \omega_k t}  g_k(x).
\label{formpsi}
\end{eqnarray}
The canonical equal-time commutation relations for the 
fundamental quantum field $\hat \psi(x,t)$
\begin{eqnarray}
\langle \left[\hat \psi(x,t),  \partial_t \hat \psi(x',t) \right] \rangle
=i \hbar \delta^D(x-x')
\end{eqnarray}
must hold. This places a further requirement that the Wronskian condition
\begin{eqnarray}
\psi_k(x,t) \partial_t \psi_k^*(x',t) 
-\partial_t \psi_k(x,t) \psi_k^*(x',t)=i \hbar g_k(x) g_k^*(x'),
\label{Wronskian}
\end{eqnarray}
together with the completeness condition
\begin{eqnarray}
\sum_k g_k(x) g_k^*(x') = \delta^D(x-x'),
\label{completeness}
\end{eqnarray}
must be satisfied. In the spatially homogeneous case 
one chooses $g_k(x)$ to be plane waves $g_k(x)= e^{ikx}$, which clearly 
satisfy (\ref{Wronskian}) and (\ref{completeness}). Once satisfied at $t=0$,
the stationarity of the scalar product also guarantees these properties 
at later times.

Given (\ref{sproduct})-(\ref{formpsi}), the purely spatial 
fields $g_k(x)$ satisfy the static eigenvalue problem
\begin{eqnarray}
\left[  - \nabla_x^2 + \chi(x) \right] g_k(x) = \omega_k^2 g_k(x).
\end{eqnarray}
The scalar product (\ref{sproduct}) can now be written in terms of $g_k$:
\begin{eqnarray}
\left( g_k(x) . g_q(x) \right)  = \int dx~  g_k^*(x) g_q(x),
\end{eqnarray} 
which is familiar from quantum mechanics. 
With these definitions orthogonality of the set $\{g_k\}$ is equivalent 
to orthogonality in $\{\psi_k\}$. 

\section{Initial conditions and renormalization}
\label{secIII}

To fully specify our initial value problem we need to choose
initial conditions for the fields. The initial fields must also 
be spatially periodic. 
For the mean field we choose a family of Gaussian profiles
\begin{eqnarray}
&& \varphi(x,t=0) = \phi_0 \exp\left[ - {x^2 \over 2 A} \right],
\label{mfield}
\end{eqnarray}
where $\phi_0, A$ are real parameters.
In addition we choose the conjugate momentum $\partial_t \varphi (x,t=0)=0$, 
for simplicity.
The choice of the set $\psi_k$ carries some arbitrarily for small $k$. 
As we have seen above, only a self-consistent set of modes 
solved in the background of (\ref{mfield}) can supply us with a 
globally static solution. 
This type of solution is  currently being 
investigated \cite{FredBoya,MotFrank} for specific profiles of $\varphi$.

The simplest possible set of mode fields satisfying all necessary 
requirements are plane waves
\begin{eqnarray}
g_k(x) = e^{i k \cdot x}.
\label{planewav}
\end{eqnarray}
The periodicity of the spatial boundary conditions enforces 
$k \rightarrow k_n = 2 \pi n/L$. For numerical work  
we take the corresponding lattice solution 
with $\omega_n = \sqrt{m^2 + 2 (1 - \cos (k_n \triangle x))/\triangle x^2}$, 
where $L=N \triangle x$ is the physical size of the system, 
$N$ its number of sites and $\triangle x$ the lattice spacing.  

Like most quantum field theories, our model as it stands is unphysical, 
as it collects infinite vacuum contributions as $\triangle x \rightarrow 0$. 
Simple power counting shows that  
$\lambda \phi^4$ in $1+1$ dimensions is superrenormalizable: 
there is a single 1-loop logarithmic ultraviolet divergence in 
the self-energy. There is also an additional quadratic divergence in 
the energy, characteristic of the free theory. 

To see this, consider the vacuum solutions to equation (\ref{psi}). 
For the purposes of renormalization we can assume $\varphi$ to be 
spatially homogeneous. We write $\chi$ as 
\begin{eqnarray}
\chi = m^2_\Lambda + 3 \lambda \varphi^2 + {3 \lambda \over \pi} 
\int_0^\Lambda dk~ \vert \psi_k \vert^2 \left[ 2 n_B(k) + 1 \right],
\end{eqnarray}
where the bare mass $m^2_\Lambda$ is understood to be cutoff dependent in a
way that will cancel the divergence resulting from the integral.

To define the spectrum of vacuum excitations, we choose $\chi$ to be the mass 
squared of the excitations, obtained from the curvature of the 
classical interaction potential at its minimum, $\chi=m^2$. 
Then, without symmetry breaking $\varphi=0$, we obtain the renormalization 
condition
\begin{eqnarray}
m^2 = m_\Lambda^2 + {3 \lambda \over \pi} 
\int_0^\Lambda dk {\hbar \over 2 \omega_k}, \label{renorm1}
\end{eqnarray}
with $\omega_k = \sqrt{k^2 + \chi}= \sqrt{k^2 +  m^2}$. This results in 
\begin{eqnarray}
&& m_\Lambda^2 = m^2    - {3 \lambda \over \pi} 
\int_0^{\Lambda} dk {\hbar \over 2 \sqrt{k^2 + m^2}}, \label{renorm2}\\
&& \chi = m^2 + 3 \lambda \varphi^2  + {3 \lambda \over \pi} 
\int_0^{\Lambda} 
dk \left\{ \vert \psi_k \vert^2 \left[ 2 n_B(k) +1 \right] 
- {\hbar \over 2 \sqrt{k^2 + m^2}} \right\}, \nonumber
\end{eqnarray}
where the last relation is valid at all times. Clearly we need to 
take $\Lambda \ll T$ to obtain results that are independent of $\Lambda$.  

The choice of plane waves for $\{g_k(x)\}$, apart from its simplicity,  
is not ideal, in the sense that the introduction of an inhomogeneous 
mean field causes the initial choice of basis to shift quickly to adapt 
to this background. This leads to a fast transient in the spatial profile 
of the mode functions $\psi_k$ over the set of long length scales 
that characterize the spatial inhomogeneity of the mean field. 
Consequently, this effect does not change the renormalization, 
which is an ultraviolet property.       
  
When computing the energy, we have, in addition to the 
logarithmic self-energy divergence, a quadratic divergence characteristic 
of the free theory.  The total energy has the form
\begin{eqnarray}
E = && {1 \over 2} \int dx~ \left\{ 
\vert \partial_t\varphi(x) \vert^2 + \vert \nabla \varphi(x) \vert^2 
+m^2 \varphi(x)^2 + {\lambda \over 2}  \varphi(x)^4 \right. \nonumber \\
&& \left. + \left( m^2+3\lambda \varphi^2+{3 \over 2} \lambda G^>(x) 
\right) G^>(x)
+\int {d{\bf k} \over 2 \pi}~ \left[ \vert \nabla \psi_k (x) \vert^2
+\vert \partial_t  \psi_k (x) \vert^2 \right] \right\}.  
\label{energy}
\end{eqnarray}
The quadratic divergence arises from the last integral in (\ref{energy}). 
To remove it we need to subtract the vacuum contribution to this term
\begin{eqnarray}
\Delta E = {\hbar \over \pi} \int^\Lambda_0 dk~  
\left[ {k^2 + \omega_k^2 \over 2 \omega_k} \right] \sim 
{\hbar \over \pi} \int^\Lambda dk~ k
\label{Ecounter}
\end{eqnarray} 
which is clearly a quadratically divergent integral with the 
ultraviolet cutoff $\Lambda$. To render the energy finite it is 
sufficient to subtract the ultraviolet behavior of (\ref{Ecounter}).
By subtracting the full integral in (\ref{Ecounter}) we impose 
the renormalization condition that the physical energy at zero temperature
is contained exclusively in the profile of the mean field $\varphi$.

We are now ready to study the non-equilibrium evolution of the quantum 
fields.

\section{The time evolution of inhomogeneous mean fields}
\label{secIV}

In this section we discuss the dynamical properties of the spatially 
inhomogeneous Hartree approximation. We solve 
Eqs.~(\ref{dyneqs}) and~(\ref{psi}), in the region $[-L/2,L/2]$,
with periodic boundary condition in space. 
We discretize the computational domain into $N+1$ points, 
with $x_j = j \triangle x$, $ j = -N/2, -N/2+1, ..., 0, 
1, 2, ..., N/2,$ where $\triangle x = L/N.$ We approximate $G^>(x,t)$ 
using $N$ complex base fields. 
A fourth-order symplectic integrator with a fixed 
time step $\triangle t$ was used to advance the solution to the final 
time $t_f$. In the results shown below, $L=128$, $N = 1024$, 
$\triangle t = 0.0025$.

As a zeroth-order check, we verified that the
total energy is conserved to within $0.001\%$, or one part per 100,000. 
We have also made a "convergence study", where, for one set of initial 
parameters, we performed the simulations using three different
spatial discretization $\triangle x$,  and found that the integral 
features of the solutions are unchanged with refinements of spatial 
resolutions. 

\begin{figure}
\begin{center}
\leavevmode
\psfig{file=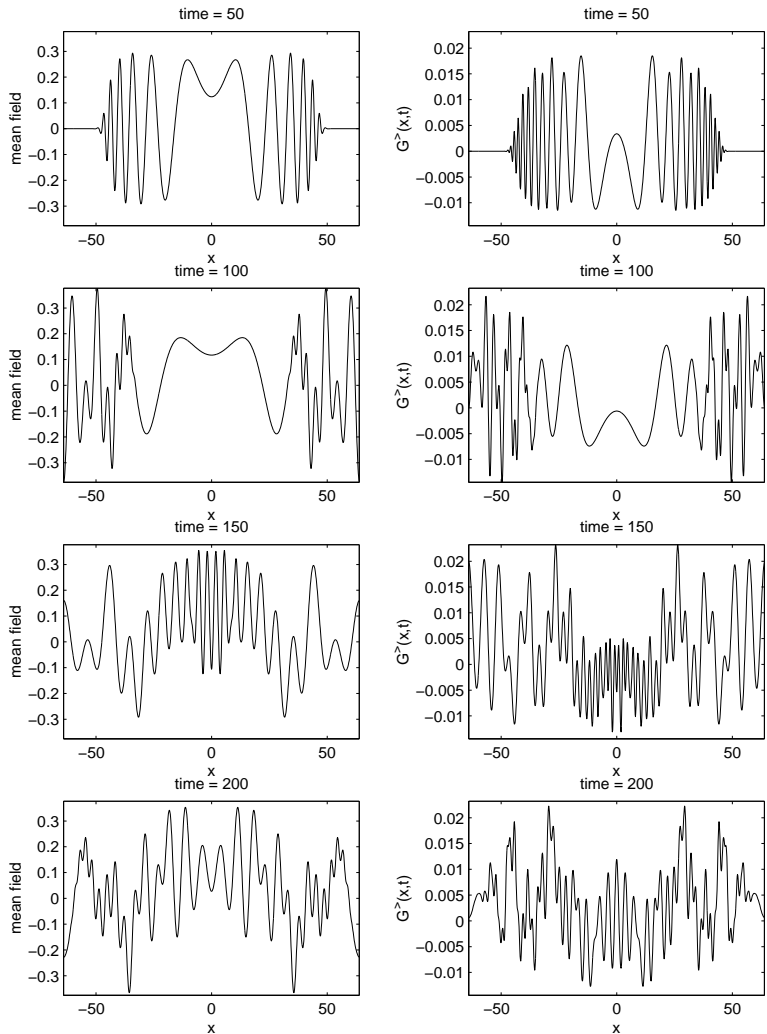,width=4.0in,silent=}
\end{center}
\caption{Snapshots of the evolution of an initial Gaussian profile 
with $\phi_0=2.0$, $A=1$ and $\lambda=0.1$, $\hbar=m^2=1$ in the mean field 
(left) and $G^>(x,t)$ (right).}
\label{fig1}
\end{figure}

In Figure~\ref{fig1} we show several snapshots of the evolution of an 
initial Gaussian profile with $\phi_0=2.0$, $A=1$, and with parameters 
$\lambda=0.1$, $\hbar=1$, $m^2=1$. 
This situation corresponds to a macroscopic local over-density 
of the field $\phi$, which could be achieved by turning on  
an external potential. At $t=0$ the external perturbation is released 
and the over-density is free to propagate and decay.
We see that the  Gaussian profile oscillates and decays 
by emitting wave packets that travel outwards from the perturbation point, 
both in its own profile, as it would happen in the purely classical case, 
and in the fluctuation modes, which correspond to particles created by the 
perturbation.

Because the fluctuations are massive, the medium is dispersive, and 
different wavelengths propagate outwards at different velocities. 
We observed that their propagation velocity is bounded by $c=1$, and that 
shorter wavelengths travel faster, as is apparent in the first few frames 
of Fig.~\ref{fig1}. Meanwhile, some of the energy of the mean field is 
transfered to the fluctuations, which qualitatively display a similar 
behavior of wave packets traveling outward from the perturbation point. 
For late times the wave packets collide due to the periodic nature of 
the spatial boundary conditions, and continue to propagate back and forth in 
the computational domain.

\begin{figure}
\begin{center}
\leavevmode
\psfig{file=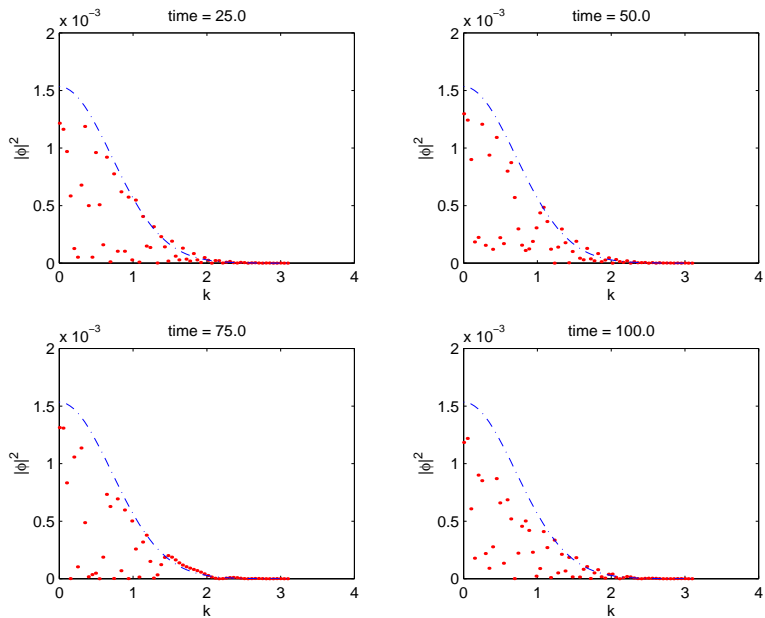,width=4.0in,silent=}
\end{center}
\caption{
Snapshots of the power spectrum of the mean field $\varphi$. The power stays 
approximately contained within the initial Gaussian envelope (dot-dashed), 
while decaying slowly as time progresses.}
\label{fig2}
\end{figure}

An important feature of the evolution is that the mean field 
does not acquire large power over small spatial scales. In fact, 
as shown in Fig.~\ref{fig2}, most of the power spectrum of the mean 
field stays contained within the initial Gaussian envelope, 
although there is some observable enhancement over smaller wavelengths  
due to the back reaction of the fluctuations. 

We now turn to the question of the long-time behavior of the dynamics.
As a preamble we should keep in mind that our approximation resumes 
interaction effects that appear at 1-loop order $\sim \lambda$. Consequently 
we can expect the approximation to be valid up to times 
$t\sim 1/\lambda=10$, for the case studies shown here\footnote{We performed 
a large variety of other evolutions, obtained by varying $\lambda$ and 
$m$ and the initial conditions for $\varphi$, with qualitatively similar 
results to those shown here at $\lambda=0.1$ and $m=1$.}. 
Collisional effects, not included in the Hartree approximation, first 
appear at two-loops ($\sim \lambda^2$) and are necessary to render the 
evolution physically valid to times of order 
$t \stackrel{>}{\sim} 1/\lambda^2$. Thus pushing the mean field 
approximation to large times requires a substantial optimism. 
Any conclusions it suggests for the properties of the 
physical system it attempts to describe in this regime 
should be met with a healthy dose of skepticism.

In any case, as we have discussed above, as time progresses 
the mean field loses some of its initial energy to the fluctuations. 
These, however, do not always carry it away 
efficiently (see also next section), leading at late times to a local 
dynamical balance between mean field and fluctuations that does not 
result in the total decay of the former.  
In this sense we see already that a state of canonical 
equilibrium is not reached, since the mean field remains non-vanishing. 
Temporal averages of local mean fields, however, will 
yield approximately zero. In this sense we may think that the system 
has reached a state of {\it microcanonical} equilibrium and that any local 
observable, averaged over time, may mimic a corresponding average obtained 
from the canonical ensemble.    

To help us decide whether we may have reached a state of microcanonical
equilibrium, we analyze the 
spectrum of fluctuations $\psi_k$.
Again, a canonical system in contact with a thermal bath would be 
expected to thermalize (in the mean field approximation) at a temperature 
consistent with equipartition. If the system thermalizes, the fluctuation 
spectrum would satisfy
\begin{eqnarray}
\langle \vert \psi_k \vert^2 \rangle = {\hbar \over 2 \omega_k} 
\left[ 2 n_B(\omega_k) +1 \right], \qquad 
\langle \vert \pi_k \vert^2 \rangle = {\hbar \omega_k \over 2 } 
\left[ 2 n_B(\omega_k) +1 \right],
\end{eqnarray}
where $\omega_k=\sqrt{k^2 +m^2(T)}$, or its form on the spatial lattice,
and $\pi_k(x,t)=\partial_t \psi_k(x,t)$.
These reduce to their classical Boltzmann forms in the limit 
of $\hbar\rightarrow 0$
\begin{eqnarray}
\langle \vert \psi_k \vert^2 \rangle = {T \over \omega^2_k}, \qquad 
\langle \vert \pi_k \vert^2 \rangle = T. 
\end{eqnarray} 
Because of the exponential decay of $n_B(\omega_k)$ for large
$\omega_k$, the vacuum contribution is typically dominant 
with $T=0$ initial conditions. We plot instead 
$\langle \vert \psi_k \vert \rangle - \langle \vert \psi_k \vert \rangle_0$, 
by subtracting out the vacuum piece.

\begin{table}
\centering
\begin{tabular}{cccc}  
\multicolumn{2}{c} {$E=0.008426$} &  
\multicolumn{2}{c} {$E=0.855148$}  \\
$A$ & $\phi_0$ &  $A$ & $\phi_0$ \\ \hline	
2.00 &  0.260128  &   2.00  &  2.60106  \\ 	
1.00 &  0.200000  &   1.00  &  2.00000  \\	
0.50 &  0.146311  &   0.50  &  1.46514  \\	
0.25 &  0.101603  &   0.20  &  1.01951  
\end{tabular}
\label{tb1}
\caption{The height $\phi_0$ and variance $A$ of
the initial Gaussian mean field $\varphi(x,0)$, 
Eq.~(\ref{mfield}) used to generate a family of evolutions with the
same energy $E$. The results for the late time fluctuation power spectrum
are shown in Figs.~\ref{fig3}, \ref{fig3b}.}
\end{table}

\begin{figure}
\centering
\leavevmode
\psfig{file=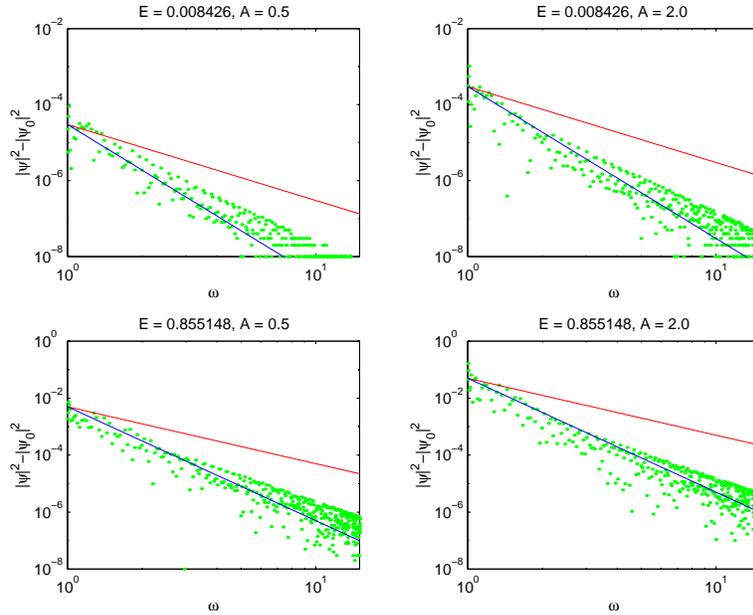,width=4.0in,silent=}
\caption{Late-time power spectra of the fluctuations 
$\vert \psi \vert^2 - \vert \psi(t=0) \vert^2$ (with subtracted 
vacuum) for various amplitudes and energies, see Table~I. 
Solid lines show $\omega^{-2}$ (upper) and $\omega^{-4}$ (lower) dependences.}
\label{fig3}
\end{figure}

\begin{figure}
\centering
\leavevmode
\psfig{file=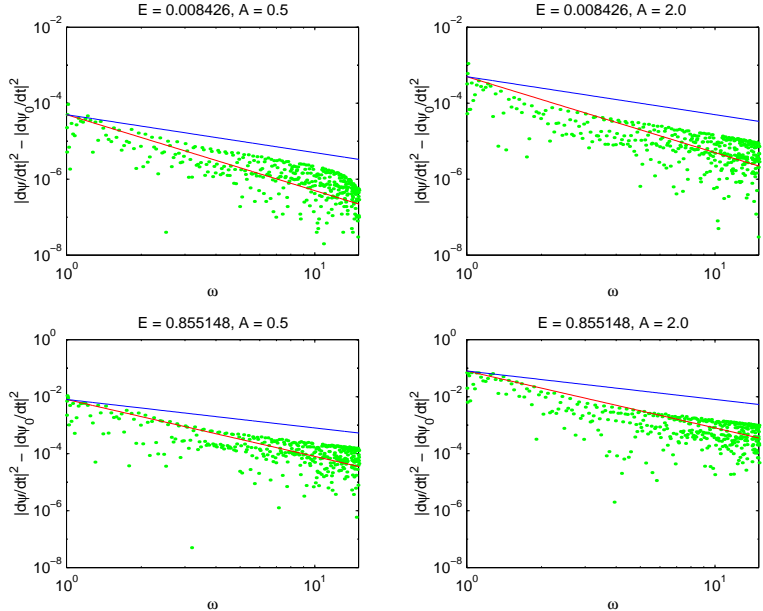,width=4.0in,silent=}
\caption{Late-time power spectra of the fluctuation conjugate momenta 
$\vert \partial_t \psi \vert^2 
- \vert \partial_t \psi(t=0) \vert^2$ (with subtracted vacuum contribution)  
for various amplitudes and energies, see Table~I. Solid lines 
show $\omega^{-1}$ (upper) and $\omega^{-2}$ (lower) dependences.}
\label{fig3b}
\end{figure}

To investigate whether the late time fluctuation power spectrum 
is in any sense universal 
we set up a family of initial conditions by varying $\phi_0$ and $A$ 
under the restriction that the renormalized total energy of the system $E$
stays constant, keeping $\lambda=0.1$ fixed.
Table~I shows the values of $\phi_0$ and $A$ used,
for two different values of the energy $E$. 
Fig.~\ref{fig3} shows several late-time fluctuation power spectra 
for initial Gaussian mean field profiles, 
while Fig.~\ref{fig3b} displays the corresponding spectra of the fluctuation
conjugate momenta.
After an initial transient, the spectra tend to settle to an approximately 
time independent profile. 
Although some of the general features of the several spectra 
are similar they differ in, for example, amplitude. 
The fluctuation spectra are reasonably fitted by a power law 
$\omega^{-\alpha}$, with $2<\alpha<4$, while that of the 
conjugate momenta may be fitted by a similar power law, but with $1<\alpha<2$.
 
These results suggest that at late times the system enters into a 
dynamical steady state with fluctuation spectra that are uncharacteristic 
of canonical thermal equilibrium, either in the classical approximation 
or for quantum fields. 
Indeed, the characterization  of the late-time spectrum as an 
approximate power law $\omega^{-\alpha}$ with $2<\alpha<4$ suggests that 
the spectrum of fluctuations is less ultraviolet dominated than in the 
classical Boltzmann approximation ($\alpha=2$) but more than in its 
quantum Bose-Einstein form. We never observed exponential suppression 
of the power spectrum at large $\omega$, as would be expected from 
quantum canonical equilibrium. 
Interestingly, this range of power laws shows that the spectrum 
of fluctuations generated dynamically by the decay of the mean field 
gives rise to expectation values with no divergence in the 
limit $k\rightarrow \infty$. If it holds in higher spatial dimensions  
this is an advantageous feature of the current 
approximation relative to purely classical field theory. Obtaining truly 
quantum occupation numbers almost certainly requires the inclusion of 
quantum scattering effects \cite{KB,Jurgens}.

The late-time statistical stationarity of the fluctuation 
power spectra suggests that in its long-time limit the 
fluctuations may decouple from the evolution.
This is what happens in {\it homogeneous} mean field theories,
where the long-time limit is characterized by $\chi(t)\rightarrow C$, 
where $C$ is a constant, dependent on the initial conditions. 
Then the modes $\psi_k (t)$  become harmonic oscillators with 
a corrected frequency. The canonical thermal equilibrium behavior 
of $\chi(x,t)$ also demands that it  becomes a constant in both 
space and time.

\begin{figure}
\begin{center}
\leavevmode
\psfig{file=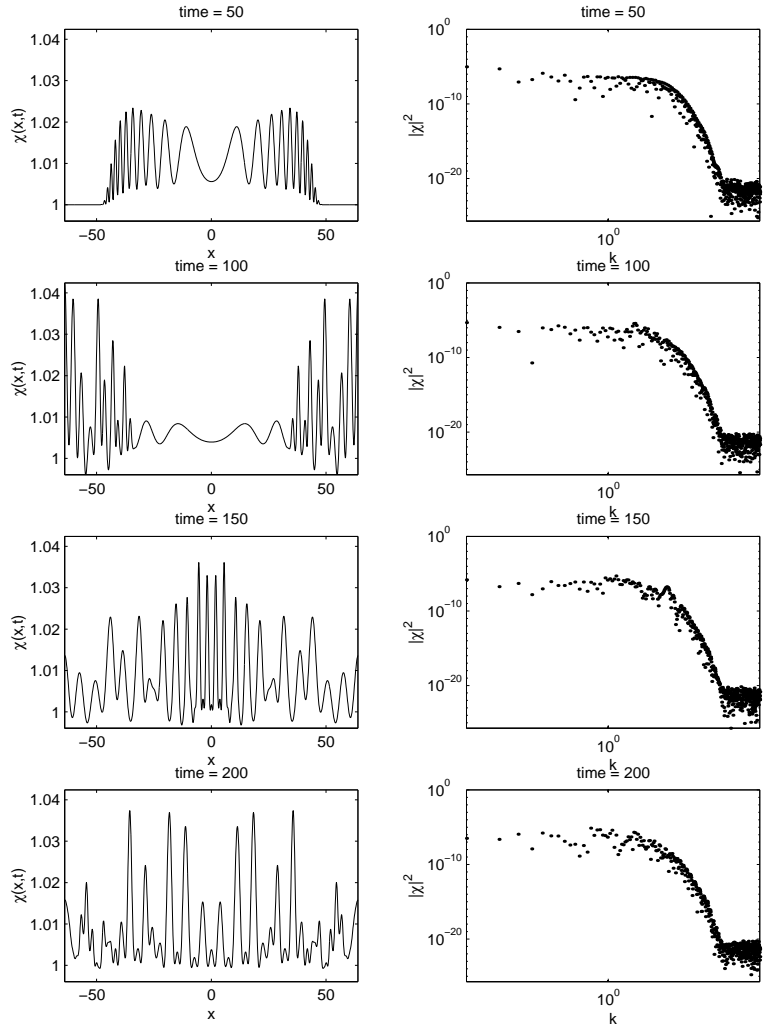,width=4.0in,silent=}
\end{center}
\caption{The evolution of $\chi(x,t)$ for an initial Gaussian mean field 
profile with $\phi_0=2.0$, $A=1$, $\lambda=0.1$, $m^2=1$ and $\hbar=1$ (left)
and of its power spectrum (right).}
\label{fig4}
\end{figure}

In order to probe this behavior, we show the dynamical evolution of 
$\chi(x,t)$ and of its power spectrum in Fig.~\ref{fig4}.
At late times the power spectrum of $\chi$ reaches approximate stationarity.
Clearly there is very little power for large $k$, showing that the modes 
in this regime are essentially decoupled. The approximate stationarity of 
the power spectrum at small $k$ suggests, however, that the system has 
reached a statistical dynamical steady-state which does not coincide with 
canonical thermal equilibrium.

It is clear that initially the mean field induces a strong perturbation 
in $\chi(x,t)$, which subsequently decays to smaller values in the 
central region. For late times $\chi(x,t)$ remains a time-dependent, 
spatially inhomogeneous field, showing that the system continues to exhibit 
complex dynamics in space-time. This dynamics, however, leads to 
no significant transport of energy among scales, as evidenced by the 
approximate stationarity of the fluctuation power spectrum.

\begin{figure}
\begin{center}
\leavevmode
\psfig{file=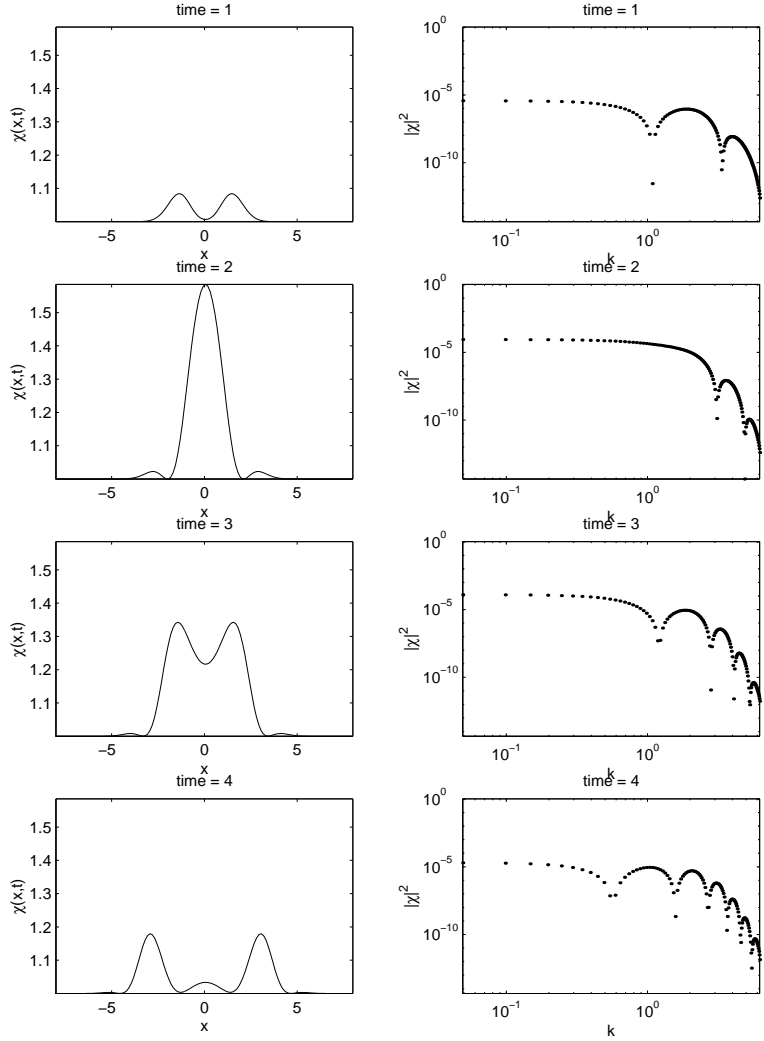,width=4.0in,silent=}
\end{center}
\caption{Early profiles of $\chi(x,t)$ (left) and corresponding 
power spectra (right) for the same parameters 
as in Fig.~\ref{fig4}. The higher $k$ modes are populated in bands that 
later coalesce to form a stable continuum spectrum.}
\label{fig5}
\end{figure}

Since most non-trivial dynamics happens at early times, we show, in 
Fig.~\ref{fig5}, the very early evolution of $\chi(x,t)$, 
together with its power spectrum.
Recall that a small $\chi$ corresponds to an approximate decoupling of the 
corresponding mode. As evidenced by Fig.~\ref{fig5}, the mean field induces 
initially significant power over a continuum of large scales. 
This suggests that the fluctuations have adapted spatially to the 
non-trivial profile of $\varphi$. The response of the high $k$ modes 
is selective and can be seen to fall into a distribution of bands. 
These bands then split into finer and finer ones to become a continuum 
at late times. 

A pattern of excited bands in the fluctuation power spectrum was also 
observed in the case of the spatially homogeneous evolution of similar 
theories \cite{Homogeneous}. Such bands are closely related to exponentially 
growing solutions of Mathieu's equation, which describes the behavior 
of an oscillator whose frequency is a periodic function, supplied in our case 
by the mean field. In the spatially inhomogeneous case, this pattern is 
complicated by the fact that the mean field possesses a continuum of 
frequencies each with a different amplitude.
To sketch these parametric resonances, we neglect the fluctuations in the 
mode equation Eq.~(\ref{psi}), since initially this contribution will be small.
Then we can write
\begin{eqnarray}
\left[ \Box +m^2 + 3 \lambda \varphi^2(x,t) \right] \psi_k (x,t) \simeq 0.
\label{psi_linear}
\end{eqnarray}
This can be written in Fourier space as 
\begin{eqnarray}
\left[ \partial_t^2 +(q^2 + m^2) \right] \psi_k (q,t) +
3 \lambda \int {d p_1 \over 2 \pi}{d p_2 \over 2 \pi} 
\varphi(p_1) \varphi(p_2) \psi_k(q-p_1-p_2,t)=0
\end{eqnarray}
The last integral describes mode mixing between the mean field
and the fluctuation field $\psi_k$. To isolate parametric instabilities
we need to further approximate it, by placing $p_1=-p_2\equiv p$, which results in
\begin{eqnarray}
\left[ \partial_t^2 +(q^2 + m^2) + 
3 \lambda \bar \varphi^2 \right] \psi_k (q,t) \simeq 0, \qquad  
\bar \varphi^2 = \int dx ~\varphi^2(x)=\int {dp \over 2 \pi}~ 
\vert \varphi(p) \vert^2 ,
\end{eqnarray} 
This equation can be mapped to Mathieu's equation, 
\begin{eqnarray}
\left[ \partial_t^2 +a + 2 \epsilon ~\cos t \right] \psi_k (q,t) = 0,
\end{eqnarray}
with $a=(q^2+m^2+3\lambda \bar \varphi^2(t=0)/2)/(2m)^2$, 
$\epsilon=3 \lambda \varphi^2(t=0)/8 m^2$. An instability analysis
of Mathieu's equation with $\bar \varphi^2(t) \simeq \bar 
\varphi^2(t=0) \cos^2 m t$ predicts  instability bands 
(exponential growing modes) for 
$a=n^2/4$, with $n$ an integer. In terms of the Fourier momentum
$q$  this  results in 
$q\simeq m \sqrt{n^2-1} = \sqrt{3}, \sqrt{8}, \sqrt{15}, ... $, 
provided $\lambda \bar \varphi^2(t=0) \ll m^2$. 

Although this analysis of parametric instabilities is necessarily 
approximate, we observed that the values for instability modes predicted 
by Mathieu's equation are realized to a good approximation in the early 
dynamics either as maxima or minima in the power spectrum of fluctuations
for $k > m$. 
For small $k$, however, the population of the power spectrum follows a broad 
continuum presumably seeded not by parametric instabilities but by terms 
in the mode mixing that act as sources. This behavior is a new feature
of the dynamics relative to the spatially homogeneous case. 
Later, more and more bands appear, resulting in a complicated behavior, 
until the spectrum reaches a continuum also for large $k$.  
These aspects of the evolution are quite complex and lie beyond the scope 
of our present analysis. They will shed 
light on the detailed mechanisms for energy flow among scales in the
model at early times, when the mean field approximation to the dynamics 
of the quantum fields is justifiable.

\section{Response around thermal equilibrium}
\label{secV}

In the previous section we showed some of the general properties
of the inhomogeneous mean field evolution starting with families of 
Gaussian profiles for the mean field and the quantum fluctuations in vacuum.
In this section, in the spirit of transport theory, we investigate the
properties of the model when perturbed around a self-consistent 
state of thermal equilibrium.

The mean field equations (\ref{dyneqs}) have as a limiting case 
a transport mean field theory characterized by a Vlasov equation.
The derivation of this transport theory is standard \cite{KB} and we 
will not repeat it here. The transport description is appropriate 
for small soft disturbances of the system around a stable 
asymptotic state, usually thermal equilibrium.
By solving the corresponding transport theory, one can estimate how 
the system responds over large spatial scales and long times 
and, in particular, how and if it returns to thermal equilibrium. 

In the context of Boltzmann transport theory, the return to thermal 
equilibrium is related to a collision kernel resulting from an 
imaginary part of the self-energy of the field in the medium.
For $\lambda \phi^4$ the lowest such self-energy is given by the 2-loop 
sunset diagram, with its imaginary part describing 
$2 \leftrightarrow 2 $ particle scattering. 
Such a self-energy is absent from our description. Thus we may 
simply expect, as is characteristic of mean field transport descriptions,
that disturbances to thermal equilibrium may never decay and that the 
response and final state of the system is described in terms of 
characteristics of its initial state (such as its energy), 
and is therefore not universal. 

As we have seen in Eq.~(\ref{Geq}), thermal initial conditions
are easily included through the definition of $G^>(x,t)$.
To ensure that we start with a self-consistent (static) solution,
we must solve the eigenvalue problem corresponding to 
the state of thermal equilibrium in the mean field approximation.
In thermal equilibrium $\varphi=0$ and the tadpole diagram acquire a 
momentum-independent mass correction $\Delta m^2(T)$.
The self-consistent solution for the modes $\psi_k$ then is 
in the form of plane waves (\ref{formpsi}), (\ref{planewav}), with frequency 
$\omega_k=\sqrt{k^2+m^2+\Delta m^2(T)}$. The mass correction 
$\Delta m^2(T)$ obeys the gap equation
\begin{eqnarray}
&& \omega_k^2 \equiv k^2 +m^2 + \Delta m^2(T), \nonumber \\
&& \Delta m^2 = {3 \hbar \lambda \over \pi} \int_0^\Lambda d k~ 
\left[ { 2 n_B(\omega_k) + 1 \over 2 \omega_k} 
- {1 \over 2 \sqrt{k^2 +m^2}}\right],
\label{gap}
\end{eqnarray}
where we have used a renormalization prescription such that 
the vacuum tadpole is fully subtracted so that the result coincides 
with (\ref{renorm1}), (\ref{renorm2}) as $T\rightarrow 0$.
We implement this solution on our spatial lattice modes $\psi_k$ 
in order to produce a static solution for the numerical evolution.
Some sample results are shown in Table~II.
\begin{table}
\centering
\begin{tabular}{cc}
$T$ & $ \Delta m^2(T)$ \\ \hline
0.0   &   0.0000  \\
0.5   &   0.0116  \\
1.0   &   0.0523  \\
2.0   &   0.1535  \\
3.0   &   0.2577  \\
5.0   &   0.4580  \\
10.0  &   0.9042  
\end{tabular}
\label{tb2}
\caption{The thermal mass correction $\Delta m^2(T)$ {\it vs.} 
the temperature $T$. The computations used a lattice propagator,
and $\lambda=0.1$, $m=1$.}
\end{table}
We verified that the solution resulting from this procedure is 
static in the absence of the mean field or an external perturbation.

An important point to note is that the form of the Bose-Einstein distribution 
is not essential for the stationarity of the solution. Equation~(\ref{gap}) 
can be satisfied for any other occupation number, provided the integral
over $k$ exists. This contrasts once again with the integrals in 
Boltzmann collisional kernels, for which the Bose-Einstein distribution 
is usually the unique solution.

We can now explore the response of the fluctuations to
a self-consistently evolving mean field or to a 
static external perturbation. 
In the examples below we will work at $T=3$, 
which is a high temperature situation since $T > m(T) = 1.12$.

\begin{figure}
\begin{center}
\leavevmode
\psfig{file=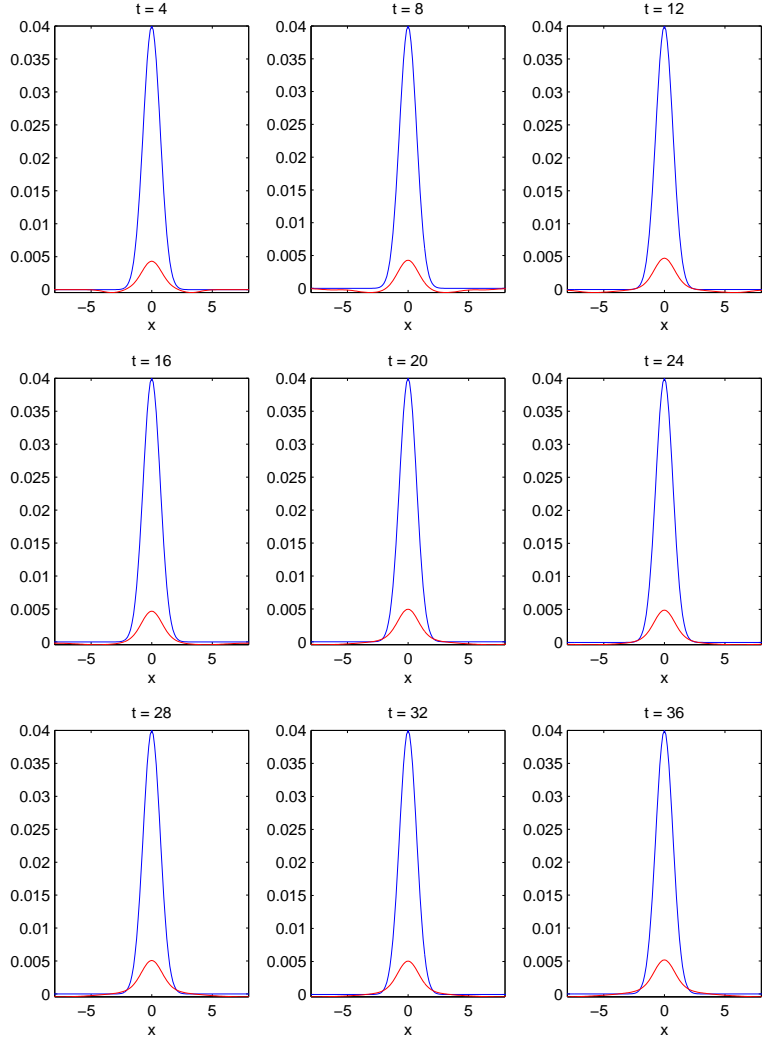,width=4.0in,silent=}
\end{center}
\caption{$G^>(x,t) - G^>_{\rm eq}$ (red) showing the response of the 
fluctuations to an externally imposed Gaussian field 
$\varphi_{\rm ext}^2$ (blue). Eq.~(\ref{mfield}), with $A=1$ and 
$\phi_0=0.2$ and at temperature $T=3$.}
\label{fig6}
\end{figure}

Consider first an external field $\varphi_{\rm ext}$. 
We assume that a static solution will be reached in this external 
field and parameterize the {\it linear} deviation from thermal equilibrium 
by generalizing the equilibrium number distribution $n_B$ to a general 
slowly varying function of space and time $n(x,k)$ such that
\begin{eqnarray}
&& n(x,k)= n_B(\omega_k) + \delta(x) {d n_B(\omega_k) \over d \omega_k}.
\end{eqnarray}  
By requiring stationarity of the solution we can conclude that 
the field $\delta(x)$ is
\begin{eqnarray}	
\delta (x) = {\varphi_{\rm ext}^2 \over \int {dk \over 2 \pi} 
{\hbar \over \omega_k} {d n_B(\omega_k) \over d \omega_k}}.
\end{eqnarray} 
We can now compare this simple expectation with the result from the 
evolution. An example is shown in Fig.\ref{fig6} for a Gaussian 
profile of the form (\ref{mfield}) with $\phi_0=0.2$ and $A=1$
  
It is striking to see that the fluctuations do not completely 
cancel out the externally imposed field, and that a globally static 
solution is not reached at long times.
Instead, the fluctuations adapt to the external potential to form a 
local bound state and some of the energy perturbation introduced at $t=0$, 
when the mean field was switched on, is carried away by wave packets
propagating outwards.

\begin{figure}
\begin{center}
\leavevmode
\psfig{file=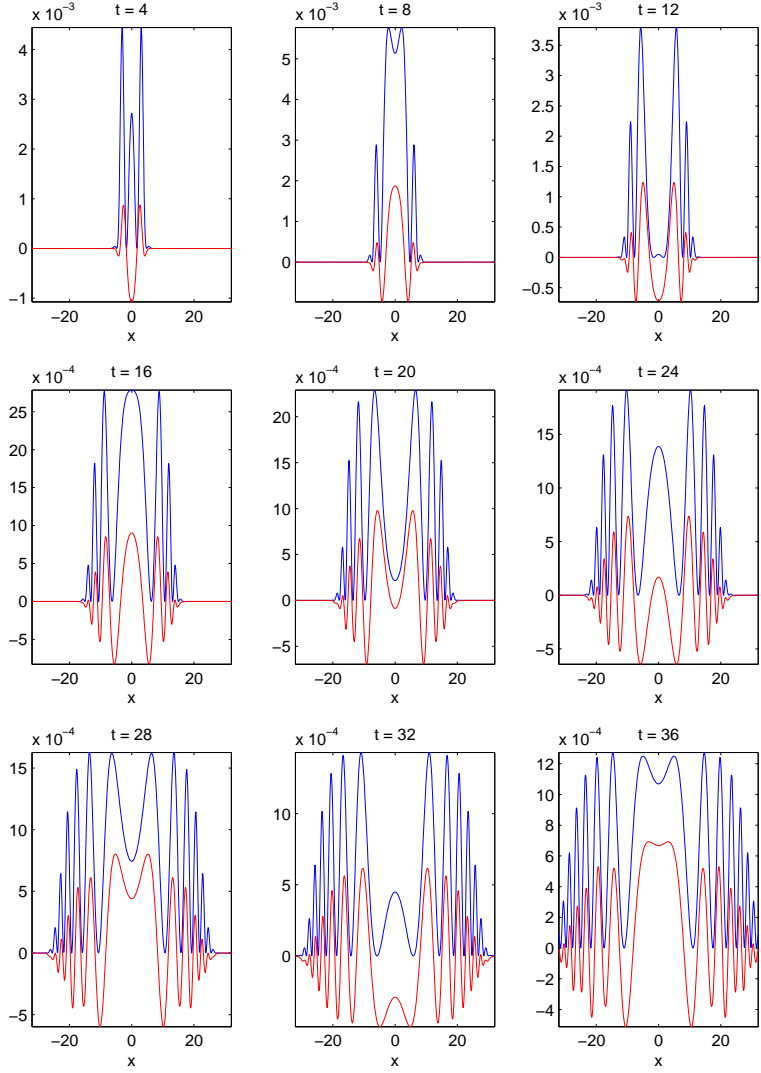,width=4.0in,silent=}
\end{center}
\caption{$G^>(x,t) - G^>_{\rm eq}$ 
(red) showing the response of the fluctuations
to a self consistently evolving mean field squared $\varphi(x,t)^2$ 
(blue) started from a Gaussian profile Eq.~(\ref{mfield}), 
with $A=1$ and $\phi_0=0.2$ and at temperature $T=3$}
\label{fig7}
\end{figure}

For the same mean field self-consistently evolving with its fluctuations 
the result is reminiscent of that of section \ref{secIV}, see Fig.\ref{fig7}. 
As in the case of the external perturbation, the self-consistently 
evolving mean field is partially shielded 
by fluctuations, but not enough to yield a globally static solution.
Instead, as the mean field evolves and its profile changes in space-time,
the fluctuations track that change on essentially the same time scale after 
an initial fast transient.

The properties of self-consistently determined fluctuations in fixed external
backgrounds both at zero and finite temperature can be naturally studied 
in our dynamical approximation. A procedure to compute minimal energy 
fluctuation modes in these potentials can be easily generated 
by introducing artificial dissipation in the evolution.

\section{Conclusions}

We have analyzed the dynamical properties of an inhomogeneous relativistic 
quantum $\lambda \phi^4$ field theory in $1+1$ dimension in the Hartree 
approximation. We considered the evolution of the fields 
starting both from a vacuum state and from a state of thermal equilibrium
at finite temperature, perturbed by a soft mean field profile.
 
The resulting evolution can be very complex, opening up many new 
possibilities in the dynamics of quantum fields arbitrarily away 
of equilibrium. Spatially inhomogeneous fields 
are necessary, {\it e.g.}, in the study of the plasma formed at 
the ultra-relativistic collision of heavy ions and its subsequent 
anisotropic cooling and in considering the evolution 
of charged quantum fields in a mean magnetic field. 

The transport of energy from the mean field to the fluctuations proceeds  
at early times selectively into intervals of spatial scales, 
which later coalesce to create a continuous spectrum, which becomes 
almost static at late times. 
This process, similarly to the spatially homogeneous case, shows 
effective time irreversibility emerging within the context of a fully 
unitary quantum field evolution. 
Although some features of the late fluctuation spectrum 
may be quite general, others, such as its amplitude, are dependent on the 
initial state. 
The dependence of the power spectrum on $\omega$ suggests that the mean 
field evolution generates occupation numbers that do not reproduce the 
quantum behavior of the exact quantum field theory but that, 
at the same time, 
are better behaved than those of the classical thermal theory. 
In particular this spectrum leads to no new ultraviolet divergence. 
It will be interesting if this characteristic holds in higher 
spatial dimensions where the classical Boltzmann distribution leads to 
ultraviolet divergent thermal contributions to expectation values of field 
correlators. Conservation of energy alone seems to suggest that such 
should be the case. 

Neither the mean field nor the spatial homogeneities in the 
fluctuations decay away completely. 
At late times the dynamics consists of propagating wave 
packets in both the mean field and fluctuations. This behavior 
may have been enhanced in the present study by the fact that there are 
no massless excitations in the spectrum and therefore there are kinematic 
thresholds prohibiting further decay. This is a general characteristic 
of the $N=1$ $\lambda \phi^4$ theory considered here.
Collectively, the characteristics of the late time evolution, 
although interesting and an improvement on the homogeneous version 
of the approximation and on the classical fields, are incompatible with a 
state of canonical thermal equilibrium. This aspect of the problem will almost certainly require the inclusion of quantum scattering effects \cite{Jurgens}.

In contrast, physical situations that are 
dominated by the behavior of (massless) Goldstone modes -such as scalar 
theories with large $N$- will behave quite differently. 
The large-$N$ limit of these 
models can also be studied for spatially inhomogeneous mean fields. Relative 
to the Hartree approximation it has the added advantage of capturing 
the correct (mean field) thermodynamics \cite{largeNSeff,Hartree1storder}. 
Unfortunately, the study of the dynamics of symmetry breaking transitions 
is most interesting in higher spatial dimensions, where, 
in order to address spatially 
inhomogeneous situations, a substantial increase in computational 
effort, relative to that of the present study, will be required.

Our results are complementary to those of Sall\'{e}, Smit and Vink 
\cite{Smit}, who studied the same model under additional averaging 
over ensembles of mean fields.
This ensemble averaging is an additional ingredient not included 
in the mean-field approximation to the dynamics of the quantum field theory 
and at face value is not allowed since the mean field already carries the 
meaning of the quantum ensemble average field expectation value. 
It may nevertheless be the means to an end -the quantum thermalization 
of the theory- at the expense of restricting the mean field to 
particular ensembles. Our results indicate that, without invoking this 
additional ensemble averaging, the mean-field dynamics does not thermalize 
for a large class of initial conditions, including the close vicinity 
of thermal equilibrium. 

Because of these shortcomings the dynamical Hartree approximation is 
best suited to study situations that are not dominated by fluctuations 
and that are not closely related to a symmetry breaking 
second order transition. 
In particular, the evolution of quasi-classical field configurations,
such as topological defects and phase interfaces in a background 
of self-consistent fluctuations, should prove very interesting.
The hydrodynamical properties of the model and,  in particular, the 
evolution of the energy momentum tensor in several situations 
are also being considered.

\section*{Acknowledgments}

We would like to thank F. Alexander,  F. Cooper, S. Habib and  
E. Mottola for useful discussions.
Numerical work was performed on LANL's SGI Origin 2000 SMP clusters 
("ASCI Blue").

\end{document}